\documentclass[a4,11pt]{article}
\usepackage{graphicx} 
\usepackage{amsmath}
\usepackage{amsfonts}
\usepackage{latexsym}
\usepackage{amssymb}
\makeatletter
 
  \@addtoreset{equation}{section}
 \makeatother

\begin{document}
\title{{\it Clean} Valuation Framework for the USD Silo\\
\normalsize \it{--An implication for the forthcoming Standard Credit Support Annex (SCSA)--}~\footnote{
This research is supported by CARF (Center for Advanced Research in Finance) and 
the global COE program ``The research and training center for new development in mathematics.''
All the contents expressed in this research are solely those of the authors and do not represent any views or 
opinions of any institutions. 
The authors are not responsible or liable in any manner for any losses and/or damages caused by the use of any contents in this research.
}}

\author{Masaaki Fujii\footnote{Graduate School of Economics, The University of Tokyo},
Akihiko Takahashi\footnote{Graduate School of Economics, The University of Tokyo}
}
\date{
First version: December 8, 2011 
}
\maketitle



\newtheorem{definition}{Definition}
\newtheorem{assumption}{$[$ A}
\newtheorem{condition}{$[$ C}
\newtheorem{lemma}{Lemma}
\newtheorem{proposition}{Proposition}
\newtheorem{theorem}{Theorem}
\newtheorem{remark}{Remark}
\newtheorem{example}{Example}
\newtheorem{corollary}{Corollary}
\def\n{{\bf n}}
\def\A{{\bf A}}
\def\B{{\bf B}}
\def\C{{\bf C}}
\def\D{{\bf D}}
\def\E{{\bf E}}
\def\F{{\bf F}}
\def\G{{\bf G}}
\def\H{{\bf H}}
\def\I{{\bf I}}
\def\J{{\bf J}}
\def\K{{\bf K}}
\def\L{{\bf L}}
\def\M{{\bf M}}
\def\N{{\bf N}}
\def\O{{\bf O}}
\def\P{{\bf P}}
\def\Q{{\bf Q}}
\def\R{{\bf R}}
\def\S{{\bf S}}
\def\T{{\bf T}}
\def\U{{\bf U}}
\def\V{{\bf V}}
\def\W{{\bf W}}
\def\X{{\bf X}}
\def\Y{{\bf Y}}
\def\Z{{\bf Z}}
\def\cala{{\cal A}}
\def\calb{{\cal B}}
\def\calc{{\cal C}}
\def\cald{{\cal D}}
\def\cale{{\cal E}}
\def\calf{{\cal F}}
\def\calg{{\cal G}}
\def\calh{{\cal H}}
\def\cali{{\cal I}}
\def\calj{{\cal J}}
\def\calk{{\cal K}}
\def\call{{\cal L}}
\def\calm{{\cal M}}
\def\caln{{\cal N}}
\def\calo{{\cal O}}
\def\calp{{\cal P}}
\def\calq{{\cal Q}}
\def\calr{{\cal R}}
\def\cals{{\cal S}}
\def\calt{{\cal T}}
\def\calu{{\cal U}}
\def\calv{{\cal V}}
\def\calw{{\cal W}}
\def\calx{{\cal X}}
\def\caly{{\cal Y}}
\def\calz{{\cal Z}}
%
\def\sskip{\hspace{0.5cm}}
\def\simleq{ \raisebox{-.7ex}{\em $\stackrel{{\textstyle <}}{\sim}$} }
\def\leqsim{ \raisebox{-.7ex}{\em $\stackrel{{\textstyle <}}{\sim}$} }
\def\ep{\epsilon}
\def\half{\frac{1}{2}}
\def\iku{\rightarrow}
\def\Iku{\Rightarrow}
\def\ikup{\rightarrow^{p}}
\def\inclusion{\hookrightarrow}
\def\cadlag{c\`adl\`ag\ }
\def\up{\uparrow}
\def\down{\downarrow}
\def\doti{\Leftrightarrow}
\def\douti{\Leftrightarrow}
\def\dochi{\Leftrightarrow}
\def\douchi{\Leftrightarrow}%
\def\yy{\\ && \nonumber \\}
\def\y{\vspace*{3mm}\\}
\def\nn{\nonumber}
\def\be{\begin{equation}}
\def\ee{\end{equation}}
\def\bea{\begin{eqnarray}}
\def\eea{\end{eqnarray}}
\def\beas{\begin{eqnarray*}}
\def\eeas{\end{eqnarray*}}
%
\def\hd{\hat{D}}
\def\hv{\hat{V}}
\def\hsd{{\hat{d}}}
\def\hx{\hat{X}}
\def\hsx{\hat{x}}
\def\bsx{\bar{x}}
\def\bsd{{\bar{d}}}
\def\bx{\bar{X}}
\def\ba{\bar{A}}
\def\bb{\bar{B}}
\def\bc{\bar{C}}
\def\bv{\bar{V}}
\def\balpha{\bar{\alpha}}
\def\bbalpha{\bar{\bar{\alpha}}}
\def\combi{\l(\begin{array}{c}\alpha\\ \beta \end{array}\r)}
\def\f{^{(1)}}
\def\s{^{(2)}}
\def\ss{^{(2)*}}
\def\l{\left}
\def\r{\right}
\def\a{\alpha}
\def\b{\beta}
\def\L{\Lambda}

\def\E{{\bf E}}
\def\P{{\bf P}}
\def\Q{{\bf Q}}
\def\R{{\bf R}}

\def\calf{{\cal F}}
\def\calp{{\cal P}}
\def\calq{{\cal Q}}

\def\ep{\epsilon}
\def\del{\delta}
\def\Del{\Delta}
\def\al{\alpha}
\def\part{\partial}
\def\ol{\overline}

\def\bull{$\bullet$}
\def\yy{\\ && \nonumber \\}
\def\y{\vspace*{3mm}\\}
\def\nn{\nonumber}
\def\be{\begin{equation}}
\def\ee{\end{equation}}
\def\bea{\begin{eqnarray}}
\def\eea{\end{eqnarray}}
\def\beas{\begin{eqnarray*}}
\def\eeas{\end{eqnarray*}}
\def\l{\left}
\def\r{\right}

\vspace{10mm}

\begin{abstract}
In the forthcoming ISDA Standard Credit Support Annex (SCSA), the trades denominated in non-G5 currencies as well as 
those include multiple currencies are expected to be allocated to the USD silo, where the contracts are collateralized by 
USD cash, or a different currency with an appropriate interest rate overlay to achieve the same economic effects.
In this paper, we have presented a simple generic valuation framework for the {\it clean price} under  the USD silo 
with the 
the detailed procedures for the initial term structure construction.
We have also shown that Cross Currency Swap (CCS) basis spread can be expressed 
as a difference between two swap rates.

\end{abstract}
\vspace{17mm}
{\bf Keywords :}
CSA, SCSA, OIS, Swap, Collateralization, basis spreads, Cross Currency Basis, Cross Currency Swaption, HJM

\newpage
\section{Introduction}
The collapse of Lehman Brothers and the ongoing EUR-zone sovereign debt crisis have brought 
the details of CSA (Credit Support Annex) into sharp focus in the pricing problems of financial contracts.
According to the survey recently carried out by KPMG~\cite{KPMG}, it seems that there is now wide recognition 
among practitioners that the appropriate discounting rate for the collateralized deals is the corresponding collateral rate.
Since the overnight rate is the standard choice for the cash collateral, it is now a market standard to use 
OIS (overnight index swap) rate to discount the future cash flows for single-currency domestic trades.

However, as we have already pointed out in the work Fujii, Shimada \& Takahashi (2010)~\cite{multiple_curves},
the choice of collateral currency does significantly affect the price of derivatives. This is because 
non-zero basis spread in CCS (cross currency swap) is nothing but the difference of the relative funding cost 
between the two currencies, which then affects the funding cost of a deal collateralized by a foreign currency.
The problem becomes even more 
complicated when there exist multiple eligible collateral currencies (or assets in general).
The right of the collateral payer to choose a collateral each time induces {\it cheapest-to-deliver option}
to the contract. As we have emphasized in Fujii \& Takahashi (2011)~\cite{collateral_choice}, the value of this option
can be quite significant when the CCS market is volatile.
Considering the ongoing concern on EUR-zone debt crisis and the historical level of CCS basis spreads,
it is easy to imagine that these complexities are evoking disputes on valuation,  and hence reducing 
market transparency and liquidity.

In order to mitigate the problem, ISDA (International Swaps and Derivatives Association) is planning to 
release SCSA (Standard Credit Support Annex)~\cite{ISDA_SCSA} in 2012. It is now argued that there will be 
one collateral requirement per currency, delivered in each currency (or converted to a single currency with an interest adjustment overlay) so that there is no embedded optionality in collateral agreements.
In SCSA, each trade will 
be assigned to one of the five DCC (Designated Collateral Currency) Silos of G5 currencies: 
USD, EUR, GBP, CHF, and JPY. 
For a trade involves only a single currency within the above G5 will be assigned to the
corresponding currency silo, and collateralized by the same currency.
In this case, the discounting rate is simply the OIS rate of the same currency. If a contracting party chooses different currency to post, the collateral rate will be adjusted in such a way that it gives the same economic effects as the standard case. Non-G5 currency as well as multiple-currency trades will be assigned to the USD silo, where the collateralization 
is done by USD cash, or a different currency with adjusted collateral rate to achieve the same economic effects.

In this note, we discuss the issues of the USD-silo valuation framework.
One possible reason that non-G5 currencies are assigned to the USD silo is  the lack of liquid domestic OIS 
markets for these currencies. If SCSA becomes the dominant choice among practitioners, 
we need to consider the fundamental issues, such as the swap curve construction, interest rates and FX modeling
under a peculiar situation, in which most of the trades are collateralized by a foreign currency (i.e. USD), and 
hence it is impossible to rely on the domestic OIS rate for valuation.
Although the theoretical background was already analyzed in our previous works, we will provide a simple closed
framework particularly for the USD silo. We have also shown that a CCS basis spread can be interpreted as a difference of
two swap rates. In the following, we will focus on the {\it clean price} of the contract
without counter party credit risk~\footnote{If there remains sizable uncollateralized exposure, there appear collateral cost adjustments as well as CVA/DVA~\cite{asymmetric_collateral}.}.

\section{Valuation under the foreign-currency collateralization}
\subsection{General discussions}
Before going to the details of the USD silo, let us review the valuation under the 
collateralization by a foreign currency. See, \cite{multiple_curves, collateral_choice} for the details.
We will assume the following conditions throughout the paper : \\
\bull Full collateralization (zero threshold) by cash.\\
\bull The collateral is adjusted continuously with zero minimum transfer amount (MTA). \\
\\
Since the daily margin call with zero threshold and zero MTA will be a part of SCSA, the above simplification will be a reasonable proxy for the reality. Now let us consider a derivative whose payout at time $T$ is given by $h^{(j)}(T)$ in terms of 
currency $j$. We suppose that currency $i$ is used as the collateral for the contract.
Note that the instantaneous return (or cost when it is negative) of holding cash collateral at time $t$ is given by
\be
y^{(i)}(t)=r^{(i)}(t)-c^{(i)}(t)
\ee
where $r^{(i)}$ and $c^{(i)}$ denote the risk-free interest rate and the collateral rate of the currency $i$, respectively.
Here, we have used the risk-free rate as the effective investment return or borrowing cost of cash after adjusting all the market and credit risks. If there remains additional funding cost independent from the credit risk, we need to include FVA (funding value adjustment)~\footnote{See for example \cite{crepey} and references therein.}. However, even in this case, it will not be categorized as the {\it clean part} since it is not agreeable between the two contracting parties, in general. 

Considering the return from the collateral account as a dividend yield, one can easily see that
\bea
h^{(j)}(t)=E_t^{Q^{(j)}}\left[e^{-\int_t^T r^{(j)}(s)ds}h^{(j)}(T)\right]
+f_x^{(j,i)}(t)E_t^{Q^{(i)}}\left[\int_t^T e^{-\int_t^s r^{(i)}(u)du}y^{(i)}(s)
\left(\frac{h^{(j)}(s)}{f_x^{(j,i)}(s)}\right)ds\right]\nn \\
\eea
where $f_x^{(j,i)}(t)$ is the FX rate at time $t$ representing the unit amount of currency $i$ in terms 
of currency $j$, and $E^{Q^{(k)}}_t[\cdot]$ denotes the time $t$ conditional expectation under the risk-neutral 
measure of currency $k$, where the money-market account of currency $k$ is used as the numeraire.
Then it can be checked that the process defined as
\bea
X(t):=e^{-\int_0^t r^{(j)}(s)ds}h^{(j)}(t)+\int_0^t e^{-\int_0^s r^{(j)}(u)du}y^{(i)}(s)h^{(j)}(s)ds
\eea
is a $Q^{(j)}$-martingale under appropriate technical conditions.
Thus, the option price process can be written as
\bea
dh^{(j)}(t)=\bigl(r^{(j)}(t)-y^{(i)}(t)\bigr)h^{(j)}(t)dt+dM^{Q^{(j)}}_t
\eea
with some $Q^{(j)}$-martingale $M^{Q^{(j)}}$.
As a result, we have the following pricing formula:
\bea
h^{(j)}(t)=E_t^{Q^{(j)}}\left[\exp\left(-\int_t^T \bigl(r^{(j)}(s)-y^{(i)}(s)\bigr)ds\right)h^{(j)}(T)\right]~.
\eea
As a corollary, when the collateral and the payment currencies are the same (i.e., currency $j$), we obtain
\bea
h^{(j)}(t)=E_t^{Q^{(j)}}\left[\exp\left(-\int_t^T c^{(j)}(s)ds\right)h^{(j)}(T)\right]~.
\eea
{\it Remark: \\In the following calculation for the USD silo, we are going to treat the currency $i$ as USD.
If there exists a domestic OIS market for currency $j$ as well as CCS of currency $(i,j)$ pair,
we can decompose the discounting rate as 
\be
r^{(j)}-y^{(i)}=c^{(j)}+y^{(j,i)}
\ee
where $y^{(j,i)}:=y^{(j)}-y^{(i)}$ and $c^{(j)}$ is the overnight (and hence collateral) rate of currency $j$.
In this case, as we have explained in \cite{multiple_curves, collateral_choice}, it is possible to construct the term structures of $c^{(j)}$ and $y^{(j,i)}$ separately by using the OIS and CCS market information.}

\subsection{Discounting under the USD silo}
In the reminder of the paper, we  fix currency $i$ as USD, and denote
\be
\overline{r}^{(j)}:=r^{(j)}-y^{(i)}
\ee
as the effective discounting rate under the USD collateralization. We define the 
USD collateralized zero coupon bond as
\be
\overline{D}^{(j)}(t,T)=E_t^{Q^{(j)}}\left[e^{-\int_t^T \overline{r}^{(j)}(s)ds}\right]~.
\ee
It is also useful to consider the USD adjusted collateral account defined by
\be
\overline{\beta}^{(j)}(t)=\exp\left(\int_0^t \overline{r}^{(j)}(s)ds\right)~,
\ee
and then it is straightforward to check that $\overline{D}^{(j)}(\cdot, T)/\overline{\beta}^{(j)}(\cdot)$
is a positive $Q^{(j)}$-martingale. Thus, we can define the USD-collateralized forward measure $Q^{(j)}_T$
by
\bea
\left.\frac{dQ^{(j)}_T}{dQ^{(j)}}\right|_t=\frac{\overline{D}^{(j)}(t,T)}{\overline{\beta}^{(j)}(t)\overline{D}^{(j)}(0,T)}~.
\eea
Using these notations, the option price in the previous section can also be written as
\be
h^{(j)}(t)=\overline{D}^{(j)}(t,T)E^{Q^{(j)}_T}_t\left[h^{(j)}(T)\right]~.
\ee

\subsection{FX forward under the USD silo}
Now, let us consider the FX forward contract between currency $j$ and $i$ (=USD) collateralized by USD.
We define the $T$-maturing FX forward at time $t$ as:\\
\bull At the current time $t$, the two parties agree to exchange the 
unit amount of currency $j$ with $K$ unit of currency $i$ at time $T$.\\
\bull The FX forward rate $f_x^{(i,j)}(t,T)$ is the amount of $K$ that makes the present value 
of the above exchange at time $t$ zero.\\

From the above description, the corresponding $K$ should satisfy
\bea
f_x^{(i,j)}(t)E^{Q^{(j)}}_t\Bigl[e^{-\int_t^T\overline{r}^{(j)}(s)ds}\bold{1}\Bigr]=E^{Q^{(i)}}_t\left[
e^{-\int_t^T c^{(i)}(s)ds}K\right]
\eea
and hence
\bea
f_x^{(i,j)}(t,T)=f_x^{(i,j)}(t)\frac{\overline{D}^{(j)}(t,T)}{D^{(i)}(t,T)}
\label{fxfwd}
\eea
where the USD zero coupon bond collateralized by the USD cash is defined by
\bea
D^{(i)}(t,T)=E^{Q^{(i)}}_t\Bigl[e^{-\int_t^T c^{(i)}(s)ds}\Bigr]~.
\eea
For the later purpose, let us define the measure $Q^{(i)}_T$ by
\be
\left.\frac{dQ^{(i)}_T}{dQ^{(i)}}\right|_t=\frac{D^{(i)}(t,T)}{\beta^{(i)}(t) D^{(i)}(0,T)}
\ee
where
\be
\beta^{(i)}(t)=\exp\left(\int_0^t c^{(i)}(s)ds\right)~.
\ee
Then, it is easy to check that $f_x^{(i,j)}(\cdot, T)$ is a $Q^{(i)}_{T}$-martingale, and hence
\bea
f_x^{(i,j)}(t,T)=E^{Q^{(i)}_T}_t\Bigl[f_x^{(i,j)}(T)\Bigr]
\eea
for $t\leq T$.

Similar calculation gives the FX forward between the two non-USD currencies $(j,k\neq i)$ under 
the USD silo as
\be
f_x^{(j,k)}(t,T)=f_x^{(j,k)}(t)\frac{\overline{D}^{(k)}(t,T)}{\overline{D}^{(j)}(t,T)}~.
\ee
Note that the FX forward does depend on the choice of collateral currency. Although we have a
currency triangle relation 
\be
f_x^{(i,j)}(t,T)f_x^{(j,k)}(t,T)=f_x^{(i,k)}(t,T)
\ee
within the USD silo, it is not true once we mix different collateral currencies, in general.
If all the multi-currency trades are allocated to the USD silo in the SCSA,
the complexity associated with the violation of currency triangle under the current CSA
will be significantly reduced once SCSA becomes the standard choice among practitioners~\footnote{
See the related discussion in \cite{dynamic_basis}.}.

\section{Fundamental instruments in the USD silo}
We assume that there exist an Fixed-versus-LIBOR interest rate swap market of currency $j$ and
a $(j,i)$-MtMCCS (Mark-to-Market Cross Currency Swap) market. Both types of contracts
are assumed to be USD collateralized. For simplicity, we will assume the LIBOR tenors are the same 
for both of the products. The methodology adopted in \cite{multiple_curves, collateral_choice} are readily 
applicable to eliminate this assumption.

\subsection{Interest rate swap (IRS) }
\label{sec_IRS}
Let us create a time grid $\{T_n\}_{n\geq 0}$ with the time span corresponding to the relevant LIBOR tenor.
At time $t=0$, the consistency condition of market quote of a $T_0$-into $T_M$-maturing interest rate swap denoted by $S_M^{(j)}$ is given by
\bea
S_M^{(j)}\sum_{m=1}^M \Del_m \overline{D}^{(j)}(0,T_m)=\sum_{m=1}^M\del_m \overline{D}^{(j)}(0,T_m)
E^{Q^{(j)}_{T_m}}\Bigl[L^{(j)}(T_{m-1},T_m)\Bigr]~.
\eea
Here, we distinguish the day-count fraction of fixed and floating legs by $\Del$ and $\del$, which are not
necessarily the same.  $L^{(j)}(T_{m-1},T_m)$ denotes the LIBOR of currency $j$ that is fixed at time $T_{m-1}$ 
with the interest accrual period of $[T_{m-1},T_m]$. Thus the market par rate is expressed as
\be
S_M^{(j)}=\frac{\sum_{m=1}^M\del_m \overline{D}^{(j)}(0,T_m)
E^{Q^{(j)}_{T_m}}\Bigl[L^{(j)}(T_{m-1},T_m)\Bigr]}{\sum_{m=1}^M \Del_m \overline{D}^{(j)}(0,T_m)}
\ee
using the LIBORs and USD-collateralized zero coupon bonds.

\subsection{Mark-to-Market Cross Currency Swap}
\label{sec_CCS}
We now consider a MtMCCS between the non-G5 currency $j$ and USD denoted by currency $i$.
The CCS consists of the two legs, one is the leg of currency $j$ and the other is of USD.
The currency-$j$ leg pays LIBOR of currency $j$ added by the fixed spread at each LIBOR 
maturing time and also contains the initial and final notional exchanges.
On the other hand, USD leg pays USD LIBOR flat with the notional exchanges. In addition, 
the USD notional is refreshed at each LIBOR fixing time so that the USD notional for the 
next accrual period is equivalent to that of the $j$-leg at the start of the period.

The present value of $j$-leg of a $T_0$-start $T_M$-maturing MtMCCS is then given by
\bea
PV_j&=&-\overline{D}^{(j)}(0,T_0)+\overline{D}^{(j)}(0,T_M) \nn \\
&&+\sum_{m=1}^M \del_m^{(j)}\overline{D}^{(j)}(0,T_m)\Bigl(E^{Q^{(j)}_{T_m}}\bigl[L^{(j)}(T_{m-1},T_m)\bigr]+B_M\Bigr)
\eea
where $B_M$ is the basis spread for the period-$M$ swap. Using the result of IRS, it can be simplified as
\bea
PV_j&=&-\overline{D}^{(j)}(0,T_0)+\overline{D}^{(j)}(0,T_M)+\Bigl(\frac{\Del}{\del}S_M^{(j)}+B_M\Bigr)\sum_{m=1}^M \del_m^{(j)}\overline{D}^{(j)}(0,T_m)~.
\eea
Here, we have approximated the ratios of day-count fraction $(\Del_m/\del_m)$ by the single constant $\Del/\del$ for 
simplicity~\footnote{For example, if the floating leg is based on 
Act360, and the fixed leg Act365, the ratio $\Del/\del$ is given by $\Del/\del=360/365$.}.

For the USD leg, the present value in term of currency $j$ is given by
\bea
PV_i&=&-\sum_{m=1}^M E^{Q^{(i)}}\left[e^{-\int_0^{T_{m-1}}c^{(i)}(s)ds}f_x^{(i,j)}(T_{m-1})\right]/f_x^{(i,j)}(0) \nn \\
&+&\sum_{m=1}^ME^{Q^{(i)}}\left[e^{-\int_0^{T_m}c^{(i)}(s)ds}f_x^{(i,j)}(T_{m-1})
\Bigl(1+\del_m^{(i)}L^{(i)}(T_{m-1},T_m)\Bigr)\right]/f_x^{(i,j)}(0)\nn 
\eea
which can be simplified as
\bea
PV_i=\sum_{m=1}^M \del_m^{(i)}D^{(i)}(0,T_m)E^{Q^{(i)}_{T_m}}
\left[f_x^{(i,j)}(T_{m-1}){\rm LOIS}^{(i)}(T_{m-1},T_m)\right]/f_x^{(i,j)}(0)
\eea
where we have defined the USD LIBOR-OIS spread by
\be
{\rm LOIS}^{(i)}(T_{m-1},T_m)=L^{(i)}(T_{m-1},T_m)-\frac{1}{\del^{(i)}_m}
\left(\frac{1}{D^{(i)}(T_{m-1},T_m)}-1\right).
\ee

In order to simplify the expression further, we assume that the covariance between LIBOR-OIS and FX
is negligibly small. Furthermore, we replace the expectation $E^{Q^{(i)}_{T_m}}[f_x^{(i,j)}(T_{m-1})]$
by $f_x^{(i,j)}(0,T_{m-1})$ by neglecting the one-period timing adjustment factor.
Then, we obtain
\bea
PV_i\simeq \sum_{m=1}^M \del_m^{(i)}D^{(i)}(0;T_{m-1},T_m)B^{(i)}(0;T_{m-1},T_m)\overline{D}^{(j)}(0,T_{m-1})
\eea
where $B$ denotes the forward LIBOR-OIS spread of USD:
\be
B^{(i)}(t;T_{m-1},T_m)=E^{Q^{(i)}_{T_m}}_t\Bigl[L^{(i)}(T_{m-1},T_m)\Bigr]-\frac{1}{\del_m^{(i)}}
\left(\frac{D^{(i)}(t,T_{m-1})}{D^{(i)}(t,T_m)}-1\right)~.
\ee
We have also defined the forward zero coupon bond of USD:
\be
D^{(i)}(t;T_{m-1},T_m):=\frac{D^{(i)}(t,T_m)}{D^{(i)}(t,T_{m-1})}~.
\ee

As a result, we obtain the par CCS basis spread (except the associated approximation errors) as
\bea
B_M=\Bigl(\overline{S}^{(j)}_M-\frac{\Del}{\del}S_M^{(j)}\Bigr)
+\frac{\sum_{m=1}^M\del_m^{(i)}\Bigl(D^{(i)}B^{(i)}\Bigr)(0;T_{m-1},T_m)\overline{D}^{(j)}(0,T_{m-1})}
{\sum_{m=1}^M \del_m^{(j)}\overline{D}^{(j)}(0,T_m)}
\label{MtMCCSLIBOR}
\eea
where 
\bea
\overline{S}_M^{(j)}:=\frac{\overline{D}^{(j)}(0,T_0)-\overline{D}^{(j)}(0,T_M)}{
\sum_{m=1}^M \del_m^{(j)} \overline{D}^{(j)}(0,T_m)}
\eea
denotes the effective swap rate associated with the discounting rate of currency $j$.

\section{Curve Construction in non-G5 Markets}
\label{sec_Curve}
Now, let us discuss the curve construction in the USD silo. 
As we have mentioned earlier, we are currently assuming that there is no swap markets
collateralized by the domestic currency. In this case, we cannot separate the curve of a domestic collateral 
rate (and hence overnight rate) $c^{(j)}$ and the curve of funding spreads $y^{(i,j)}$, which is known to be straightforward in the case of G5 markets. However, as we will see shortly, it is still possible to 
extract all the necessarily forwards and discounting rates in the USD silo.

For USD domestic market, we can readily derive $\{D^{(i)}(0,T_m),~B^{(i)}(0,T_{m-1},T_m)\}$ at every relevant point of 
time by applying the method explained in \cite{multiple_curves, collateral_choice}. 
Suppose we have completed the curve construction for these USD instruments.
Then, from Eq.~(\ref{MtMCCSLIBOR}) of MtMCCS, we can see that the only unknowns are the set of $\{\overline{D}^{(j)}(0,T_m)\}$, and hence we can bootstrap them using the market quotes of IRS $\{S_M^{(j)}\}$
and MtMCCS $\{B_M\}$~\footnote{In realities, since  only a limited number of quotes are available in the market, it is impossible to carry out bootstrapping literary and we need an appropriate spline technique.}.
Suppose now that we have obtained $\overline{D}^{(j)}(0,T_{m})$ up to $m=M-1$. Then, Eq.~(\ref{MtMCCSLIBOR})
gives
\bea
\overline{D}^{(j)}(0,T_M)&=&\left\{\overline{D}^{(j)}(0,T_0)-\left(\frac{\Del}{\del}S_M^{(j)}+B_M\right)\sum_{m=1}^{M-1}\del_m^{(j)}
\overline{D}^{(j)}(0,T_m)\right.\nn \\
&&\hspace{-20mm}\left.+\sum_{m=1}^M\del_m^{(i)}\frac{D^{(i)}(0,T_m)}{D^{(i)}(0,T_{m-1})}B^{(i)}(0;T_{m-1},T_m)\overline{D}^{(j)}(0,T_{m-1})\right\}/\Bigl(1+\Del_M^{(j)} S_M^{(j)}+\del_M^{(j)}B_M\Bigr)\nn\\
\eea
which fixes $\overline{D}^{(j)}(0,T_M)$. Repeated use of the above calculation recursively determines all the relevant $\{\overline{D}^{(j)}(0,T_m)\}$.

Once we obtain the effective discounting factors of currency $j$, it is straightforward to obtain the associated 
LIBOR forwards. As before, let us suppose we have fixed $\{E^{Q^{(j)}_{T_m}}\bigl[L^{(j)}(T_{m-1},T_m)\bigr]\}$
up to $m=M-1$. Then, it is easy to see that the next forward is determined by
\bea
&&E^{Q^{(j)}_{T_M}}\bigl[L^{(j)}(T_{M-1},T_M)\bigr]=\Bigl\{
\sum_{m=1}^{M}\Del_m^{(j)}\overline{D}^{(j)}(0,T_m)S_M^{(j)} \nn \\
&&\hspace{20mm}-\sum_{m=1}^{M-1}\del_m^{(j)}\overline{D}^{(j)}(0,T_{m-1})E^{Q^{(j)}_{T_m}}\bigl[
L^{(j)}(T_{m-1},T_m)\bigr]\Bigr\}/\Bigl(\del_M^{(j)}\overline{D}^{(j)}(0,T_{M})\Bigr)~.\nn\\
\eea

\section{A Possible Adjustment for non-USD Collateral}
\label{overlay}
In SCSA, non-G5 currencies will be assigned to the USD silo and  collateralized by the USD cash.
However, if USD is the unique eligible collateral, it may induce liquidity squeeze and raise the 
USD funding cost under stressed market conditions. In order to mitigate the problem, it is being considered to accept also a 
different currency with an appropriate adjustment mechanism that guarantees the same economic effects 
as the standard case of USD collateralization.
Although we do not know the actual mechanism to be proposed in SCSA, it is easy to come up with 
one possible solution from the result of the previous sections.

If the option $h^{(j)}$ is collateralized by the USD cash, we know that its value is given by
\bea
h^{(j)}(t)=E^{Q^{(j)}}_t\left[\exp\left(-\int_t^T\overline{r}^{(j)}ds\right)h^{(j)}(T)\right]~.
\eea
where $\overline{r}^{(j)}=r^{(j)}-y^{(i)}$.
Suppose now that the same option is collateralized by the currency $j$ with the adjusted collateral rate
$\tilde{c}^{(j)}$ different from the domestic overnight rate. In this case, one obtains the option value as
\bea
\tilde{h}^{(j)}(t)=E^{Q^{(j)}}_t\left[\exp\left(-\int_t^T \tilde{c}^{(j)}(s)ds\right)h^{(j)}(T)\right]~.
\eea
Thus, if the adjusted collateral rate satisfies
\be
\tilde{c}^{(j)}(t)=\overline{r}^{(j)}(t)
\ee
we can recover the same result.

Thus, the only problem is to find a simple way to calculate such $\tilde{c}^{(j)}$
from the available market data. If we assume the existence of short-term FX swaps, 
we can observe short-term FX forward points in the market.  
Consider, for example, a spot-start (spot+one business date)-maturing FX swap between the currency $j$ and USD.
Since we know the USD overnight rate (Fed-fund rate) for the corresponding period, we can extract
$\overline{r}^{(j)}$ for the same period from the market FX forward point.
This is because the USD collateralized FX-forward is given by Eq.~(\ref{fxfwd}).
The proper choice of FX swap will depend on the details of settlement schedules of the 
relevant collateral account to be modeled. For consistency, these FX swaps should be continuously collateralized by USD,
but in the market, a very short FX swap  may not have margin calls.
However, since it involves the initial notional exchange, it seems as if one party obtains different currency by
posting the equivalent amount of USD, which makes short FX swaps approximately USD 
collateralized~\footnote{We can consider that one-time collateral rate payment is included in the final notional (which is adjusted by the FX forward points) exchange.}.

Let us consider what should be done if another non-G5 currency $k$ is posted as collateral 
with the collateral rate $\tilde{c}^{(k)}$ for the same option.
Under this situation, the option value can be expressed as
\bea
\tilde{h}^{(j)}(t)=E_t^{Q^{(j)}}\left[\exp\left(-\int_t^T \bigl(r^{(j)}(s)-\tilde{y}^{(k)}(s)\bigr)ds
\right)h^{(j)}(T)\right]
\eea
where 
\be
\tilde{y}^{(k)}(t)=r^{(k)}(t)-\tilde{c}^{(k)}(t)~.
\ee
Hence, in order to recover the same price as the standard USD collateralization, the collateral rate for 
the currency $k$ should satisfy $\tilde{y}^{(k)}=y^{(i)}$, i.e., 
\be
\tilde{c}^{(k)}(t)=r^{(k)}(t)-y^{(i)}=\overline{r}^{(k)}
\ee
but this is nothing but the effective discounting rate of currency $k$ under USD collateralization.
Therefore, it can be extracted by using the market quotes USD and currency $k$ FX swaps following the 
previous explanation.
\\\\
{\it Remark:
As we have seen, the existence of liquid short-term 
(such as tomorrow-next and spot-next) FX swap markets will be enough for the 
daily adjustment of collateral fee. However, in order to guarantee the transparency of 
general valuation of derivatives, we also need long-term cross currency swaps to derive forward expectations of 
the relevant rates.
Replacing FX swaps by repo transactions, it is easy to understand that the method explained above can be extended to non-cash collaterals, too.}

\section{Developed OIS markets in the USD silo}
At the moment, OIS still lacks liquidity relative to the standard IRS. However, it is reasonable to expect
that it will be more liquid as the increase of the use of  cash collateral and associated standard CSA
in OTC and CCP markets. In many non-G5 markets, even in some of the G5, 
advanced foreign financial firms are playing an important role as dominant suppliers of derivatives. 
For these players, particularly after SCSA gains popularity, managing the market risks of 
various currencies under USD collateralization will be one of their 
central concerns.
In addition, the recognition of non-negligible credit risk even for the banks in the LIBOR panel now calls
the usefulness of LIBOR as the reference index for the wide market participants into serious question. Considering the difficulty to separately manage its embedded risks on top of those 
associated with collateral rates, we think that it is not unrealistic that OIS (rather than IRS) will gain much 
higher liquidity in coming years.
Based on these considerations, 
we think it is worthwhile studying the possible future developments of USD-collateralized OIS and 
MtMCCOIS markets~\footnote{In fact, it seems possible to trade these products already in the current OTC market.} 
and their implications.

\subsection{USD-collateralized OIS}
The valuation of this product is done by the simple replacement of the LIBOR  in the standard IRS explained in 
Sec.~\ref{sec_IRS} by 
the compounded overnight rate, $L_{ois}$.
It is given by
\be
L_{ois}^{(j)}(T_{m-1},T_m)=\frac{1}{\del^{(j)}_m}\left(\exp\left(\int_{T_{m-1}}^{T_m}c^{(j)}(s)ds\right)-1\right)
\ee
and the market OIS rate $S^{ois}_M$ for M-period swap at $t=0$ should satisfy
\bea
S_M^{ois(j)}\sum_{m=1}^M \del_m \overline{D}^{(j)}(0,T_m)=\sum_{m=1}^M\del_m \overline{D}^{(j)}(0,T_m)
E^{Q^{(j)}_{T_m}}\Bigl[L_{ois}^{(j)}(T_{m-1},T_m)\Bigr]~,
\eea
where we have assumed, for simplicity,  that the fixed leg has the same day-count conventions as the floating leg.

The forward OIS rate at time $t$ of $T_S$-start $T_M$-maturing swap is given by
\be
S^{ois(j)}(t;T_S,T_M):=\frac{\sum_{m=S+1}^M\del_m^{(j)} \overline{D}^{(j)}(t,T_m)E_t^{Q^{(j)}_{T_m}}\Bigl[L_{ois}^{(j)}(T_{m-1},T_m)\Bigr]}{\sum_{m=S+1}^M \del_m^{(j)} \overline{D}^{(j)}(t,T_m)}
\label{fwd_OIS}
\ee
and the price of an associated payer swaption with strike $K$ is
\be
{\rm Payer}=A(0)E^{Q^{(j)}_A}\left[\Bigl(S^{ois(j)}(T_S;T_S,T_M)-K\Bigr)^+\right]
\ee
where $A$ denotes 
\be
A(t)=\sum_{m=S+1}^M \del_m^{(j)} \overline{D}^{(j)}(t,T_m)~.
\label{annuity}
\ee
Here, the annuity measure $Q^{(j)}_A$ is defined by the density:
\be
\left.\frac{dQ^{(j)}_A}{dQ^{(j)}}\right|_t=\frac{A(t)}{\overline{\beta}^{(j)}(t)A(0)}~.
\ee
\subsection{USD-collateralized MtMCCOIS}
We now consider MtMCCOIS, which is the MtMCCS but the LIBORs of the two currencies are 
replaced by the compounded OIS rates.
As in Sec.~\ref{sec_CCS}, let us consider an M-period swap between USD and currency $j$.
The present value of currency $i$ (or USD) turns out to be significantly simple.
Since it pays USD overnight rate and also collateralized by USD, the value becomes zero:
\be
PV_i=0~.
\ee
This can be seen by the expression
\bea
E_t^{Q^{(i)}_{T_m}}\Bigl[L^{(i)}_{ois}(T_{m-1},T_m)\Bigr]=\frac{1}{\del_m^{(i)}}\left(
\frac{D^{(i)}(t,T_{m-1})}{D^{(i)}(t,T_m)}-1\right)
\eea
which leads to vanishing $B^{(i)}(t,T_{m-1},T_m)$ for any $t\leq T_{m-1}$.

On the other hand, the value of $j$ leg is given by
\bea
PV_j&=&-\overline{D}^{(j)}(0,T_0)+\overline{D}^{(j)}(0,T_M)+\Bigl(S_M^{ois(j)}+B^{ois}_M\Bigr) \sum_{m=1}^M \del_m^{(j)}\overline{D}^{(j)}(0,T_m)
\eea
where $B_M^{ois}$ is the MtMCCOIS spread for the M-period swap. Since $PV_i=0$, the 
par basis spread should satisfy $PV_j=0$ and hence
\bea
B_M^{ois}=\overline{S}_M^{(j)}-S_M^{ois(j)}
\eea
which is the swap spread between the effective discounting rate and OIS in the USD-silo.
One can see that the bootstrapping the $\overline{D}^{(j)}$ is significantly simpler than that of Sec.~\ref{sec_Curve}:
\bea
\overline{D}^{(j)}(0,T_M)&=&\frac{\overline{D}^{(j)}(0,T_0)-\left(S_M^{ois(j)}+B_M^{ois}\right)\sum_{m=1}^{M-1}\del_m^{(j)}
\overline{D}^{(j)}(0,T_m)}{1+\del_M^{(j)}(S_M^{ois(j)}+B_M^{ois})}~.\nn\\
\eea

It is also straightforward to consider a forward MtMCCOIS swap. If we consider a $T_S$-start, $T_M$-maturing
forward swap, then the forward MtMCCOIS basis spread at time $t$ is 
\bea
B^{(ois)}(t;T_S,T_M):=\overline{S}^{(j)}(t;T_S,T_M)-S^{ois(j)}(t;T_S,T_M)~,
\eea
where
\bea
\overline{S}^{(j)}(t;T_S,T_M)=\frac{\overline{D}^{(j)}(t,T_S)-\overline{D}^{(j)}(t,T_M)}{
\sum_{m=S+1}^M \del_m^{(j)}\overline{D}^{(j)}(t,T_m)}
\eea
and $S^{ois(j)}(t;T_S,T_M)$ is given in Eq.~(\ref{fwd_OIS}).

\section{Heath-Jarrow-Morton Framework in the USD Silo}
\label{sec_HJM}
Is is straightforward to assign HJM framework for the USD silo term structures.
Firstly, let us define the instantaneous forward rate for the USD collateralized discounting rate:
\bea
e^{-\int_t^T \overline{f}^{(j)}(t,s)ds}=E^{Q^{(j)}}_t\left[e^{-\int_t^T \overline{r}^{(j)}(s)ds}\right]
\eea
or equivalently
\bea
\overline{f}^{(j)}(t,T)=-\frac{\partial}{\partial T}\ln \overline{D}^{(j)}(t,T)~.
\eea
Note that, this rate can be negative when the CCS basis spread is significantly wide and negative.
By following the same arguments in the previous work~\cite{dynamic_basis}, we can see its dynamics
can be written in the following form in general:
\bea
d\overline{f}^{(j)}(t,s)=\sigma^{(j)}(t,s)\cdot \left(\int_t^s \sigma^{(j)}(t,u)du\right)dt
+\sigma^{(j)}(t,s)\cdot dW_t^{Q^{(j)}}~,
\eea
where $W^{Q^{(j)}}\in \mathbb{R}^d$ is the $d$-dimensional $Q^{(j)}$ Brownian motion, and 
$\sigma^{(j)}(\cdot,s)\in \mathbb{R}^d$ is a $d$-dimensional process adapted to $\mathbb{F}$,
which is the augmented filtration generated by $W^{Q^{(j)}}$.  The $(\cdot)$ denotes the 
summation on the $d$-dimensional indexes.

Now, let us consider the forward LIBOR 
\be
L_m^{(j)}(t):=E^{Q^{(j)}_{T_m}}_t\Bigl[L^{(j)}(T_{m-1},T_m)\Bigr]~.
\ee
Since it should be a $Q^{(j)}_{T_m}$-martingale, it can be expressed as
\bea
dL^{(j)}_m(t)=\Sigma_{m}^{(j)}(t)\cdot\left(\int_t^{T_m}\sigma^{(j)}(t,u)du\right)dt+\Sigma_{m}^{(j)}(t)\cdot dW_t^{Q^{(j)}}~. 
\eea
where, $\Sigma_{m}^{(j)}(\cdot)\in \mathbb{R}^{d}$ is also an $\mathbb{F}$-adapted process. Note that 
the Brownian motion under the USD collateralized forward measure $Q^{(j)}_{T_m}$ is given by
\bea
dW_t^{Q^{(j)}_{T_m}}=dW_t^{Q^{(j)}}+\left(\int_t^{T_m}\sigma^{(j)}(t,u)du\right)dt~.
\eea
If one is interested in the OIS market in the USD silo, we can simply replace the forward LIBOR 
by that of the compounded overnight rate
\be
L_m^{ois(j)}(t):=E_t^{Q^{(j)}_{T_m}}\Bigl[L^{(j)}_{ois}(T_{m-1},T_m)\Bigr]~
\ee
and assign its arbitrage-fee dynamics by
\bea
dL^{ois(j)}_m(t)=\sigma_m^{(j)}(t)\cdot\left(\int_t^{T_m}\sigma^{(j)}(t,u)du\right)dt+\sigma_m^{(j)}(t)\cdot dW_t^{Q^{(j)}}~,
\eea
where $\sigma_m(\cdot)\in \mathbb{R}^d$ is adapted to $\mathbb{F}$~\footnote{If one needs to model 
LIBOR and OIS at the same time, it would be better to model LIBOR-OIS spread dynamics instead of LIBOR itself, which 
makes it is easier to guarantee the positivity of the spread. See the related discussion in \cite{dynamic_basis}.}.

For the FX modeling, it is 
easy to check that the arbitrage free dynamics of the spot FX of currency $j$ relative to USD (or $i$) can be specified as
\bea
df_x^{(j,i)}(t)/f_x^{(j,i)}(t)=\bigl(\overline{r}^{(j)}(t)-c^{(i)}(t)\bigr)dt+\sigma_X^{(j,i)}(t)\cdot dW_t^{Q^{(j)}}~,
\eea
where $\sigma_X^{(j,i)}(t)\in \mathbb{R}^d$ is an $\mathbb{F}$-adapted process as the volatility of FX rate.
Similarly, the FX rate between the two non-G5 currencies $(j,k)$ in the USD silo can be expressed as
\bea
df_x^{(j,k)}(t)/f_x^{(j,k)}(t)=\bigl(\overline{r}^{(j)}(t)-\overline{r}^{(k)}(t)\bigr)dt+
\sigma_X^{(j,k)}(t)\cdot dW_t^{Q^{(j)}}~.
\eea

The above set of formulas provide the generic framework for the term structure modeling in the USD silo.
For the concrete implementation, we need to discretize the forward rates appropriately,
which will result in a generalization of popular BGM-like model. Pricing of various options and calibration 
procedures under the new framework will be our future research topics.

\section{Swaption on MtMCCOIS in the USD silo}
For financial firms, especially those operating in non-G5 markets allocated to the USD silo,
the risk management of USD funding cost has a critical importance. In order to hedge the associated 
future uncertainties, the existence of options on cross currency swap is highly
desirable, since it is actually the main funding source of USD in the market.
Thus, in this section, we consider the valuation problem of swaptions of MtMCCOIS in the USD silo.

Let us consider $T_S$-expiring basis payer option with strike $K$ on the $T_S$-start, $T_M$-maturing swap mentioned in the 
previous section. In this case, the present value of the swaption can be expressed by
\bea
{\rm Payer}^{ccs}&=&A(0)E^{Q^{(j)}}\left[\Bigl(B^{ois(j)}(T_S;T_S,T_M)-K\Bigr)^+\right]\nn \\
&=&A(0)E^{Q^{(j)}_A}\left[\Bigl(\overline{S}^{(j)}(T_S;T_S,T_M)-
S^{ois(j)}(T_S;T_S,T_M)-K\Bigr)^+\right]
\eea
where $A$ is the corresponding annuity defined in Eq.~(\ref{annuity}).
It is important to note that both of the $\overline{S}^{(j)}(\cdot;T_S,T_M)$ and 
$S^{ois(j)}(\cdot;T_S,T_M)$ are $Q^{(j)}_A$-martingales, and the problem is equivalent to 
the valuation of a spread option.

Now, let us consider the SDEs for the two swap rates. In the following, we use the 
simplified notations 
\bea
S_1(t)&:=&\overline{S}^{(j)}(t;T_S,T_M)\\
S_2(t)&:=&S^{ois(j)}(t;T_S,T_M)
\eea
and  also omit the currency specification of $j$.
It is not difficult once we remember that they are martingales under the annuity measure.
Tedious but straightforward calculation gives you
\bea
&&dS_1(t)=\sigma_1(t)\cdot dW_t^{Q^A}\nn\\
&&\quad:=\left[\frac{\overline{D}(t,T_M)}{A(t)}\int_{T_S}^{T_M}
\sigma(t,s)ds+
\frac{S_1(t)}{A(t)}\sum_{m=S+1}^M\del_m\overline{D}(t,T_m)\int_{T_S}^{T_m}\sigma(t,s)ds
\right]\cdot dW_t^{Q_A}~, \nn\\
\eea
and 
\bea
&&dS_2(t)=\sigma_2(t)\cdot dW_t^{Q^A}\nn \\
&&\quad:=\left[\frac{1}{A(t)}\sum_{m=S+1}^M \del_m \overline{D}(t,T_m)
\left\{\sigma_m(t)+\Bigl(S_2(t)-L^{ois}_m(t)\Bigr)\int_{T_S}^{T_m}\sigma(t,s)ds\right\}\right]\cdot dW_t^{Q_A}~\nn\\
\eea
where $W^{Q_A}\in \mathbb{R}^d$ is the $d$-dimensional Brownian motion under the annuity measure $Q_A$. 

Because of the complicated state dependency in their volatility terms, it is impossible to derive a closed form 
solution for CCS swaption. However, once we specify all the details of HJM model in Sec.~\ref{sec_HJM},
one can obtain the approximate solution by using asymptotic expansion technique ( see, for example, \cite{T,KT,TTT} and references therein.).
The actual derivation under the realistic model specifications will be studied elsewhere~\cite{future_work}.
\\
\\
{\it Remark: Once the CCS swaption market is developed, it can be used to calibrate volatility and correlation 
parameters of $\overline{f}$ for non-USD currencies. Alternatively, based on the similar mechanism discussed in 
Sec.~\ref{overlay}, it is possible to setup the swap market for the effective discounting rate $\overline{r}$ directly,
whose par swap rate is given by $\overline{S}_M$. If there exists a swaption market on this new instrument, it 
can play the same role as that of CCS but in a slightly simpler way.}

\section{Conclusion}
In the forthcoming SCSA, non-G5 currencies and multi-currency trades are expected to be  allocated to the USD silo, 
where the contracts are collateralized by USD or other currencies with appropriate collateral rate adjustments.
In this paper, we have presented a simple and generic valuation framework for the {\it clean price} under the USD silo, 
with detailed explanation for the initial term structure construction. We have also discussed the possible interest rate overlay 
method to make the non-USD collateralized trade economically equivalent to those with the standard USD collateralization.
We found that the Cross Currency Basis 
can be expressed as the difference of the two swap rates and discussed the pricing method of its swaption.


\end{document}